\documentclass[manuscript]{aastex}

\usepackage{graphics}
\usepackage{amsmath,amssymb}

\shorttitle{Discovery of Diffuse Hard X-ray Emission around Jupiter with {\it Suzaku}}
\shortauthors{Ezoe et al.}

\begin{document}

\title{Discovery of Diffuse Hard X-ray Emission around Jupiter \\with {\it Suzaku}}

\author{Y. Ezoe, K. Ishikawa and T. Ohashi}
\affil{Department of Physics, Tokyo Metropolitan University, 1-1 Minami-Osawa, Hachioji,
 Tokyo 192-0397, JAPAN}
\email{ezoe@phys.metro-u.ac.jp}

\author{Y. Miyoshi} 
\affil{Solar-Terrestrial Environment Laboratory, Nagoya University, Furo-cho, Chikusa-ku,
Nagoya 464-8601, JAPAN}

\author{N. Terada} 
\affil{Department of Geophysics, Tohoku University, 6-3 Aoba, Aramaki, Aoba-ku, Sendai, 
Miyagi 980-8578, JAPAN}

\author{Y. Uchiyama} 
\affil{Stanford Linear Accelerator Center, Standford University, 2575 Sand Hill Road,
Menlo Park, CA 94025, USA}

\and

\author{H. Negoro} 
\affil{Department of Physics, Nihon University, 1-8-14 Kanda Surugadai, Chiyoda,
Tokyo 101-8308, JAPAN}

\begin{abstract}
We report the discovery of diffuse hard (1--5 keV) X-ray emission 
around Jupiter in a deep 160 ks {\it Suzaku} XIS data.
The emission is distributed over $\sim16\times8$ Jovian radius 
and spatially associated with the radiation belts and the Io Plasma Torus.
It shows a flat power-law spectrum with a photon index of 1.4$\pm$0.2
with the 1--5 keV X-ray luminosity of (3.3$\pm$0.5)$\times$10$^{15}$ erg s$^{-1}$. 
We discussed its origin and concluded that it seems to be truly diffuse,
although a possibility of multiple background point sources can not be
completely rejected with a limited angular resolution.
If it is diffuse, the flat continuum indicates that X-rays arise by the 
non-thermal electrons in the radiation belts and/or the Io Plasma Torus. 
The synchrotron and bremsstrahlung models can be rejected from the necessary 
electron energy and X-ray spectral shape, respectively. The inverse-Compton 
scattering off solar photons by ultra-relativistic (several tens MeV) electrons 
can explain the energy and the spectrum but the necessary electron density is 
$\gtrsim$10 times larger than the value estimated from the empirical model of 
Jovian charge particles. 
\end{abstract}

\keywords{planets and satellites: individual (Jupiter, Io) --- X-rays: general}

\section{Introduction}

In situ measurements and ground-based observations have shown the existence 
of Jupiter's high-energy electron Van Allen radiation belts with evidence
for electrons at energies of keV to 50 MeV \cite{ber76,van75,fis96,bol02}. 
Theoretically, electrons are energized by the betatron acceleration via the 
inward radial diffusion \cite{goe79,miy99} and recently non-adiabatic acceleration 
via wave-particle interactions has been discussed \cite{hor08}.  
The trapped electrons lose their energy through synchrotron radio 
emission and some fraction are absorbed by Jovian moons and rings \cite{bag04}. 
However, there remains much debate over the precise energy spectrum and 
spatial distribution of the energetic electrons in the radiation belts. 

Jupiter is the most luminous planet in the solar system at X-ray wavelength 
\cite{met83,wai97,bha07}. 
It emits X-rays from magnetic poles by the charge exchange interaction 
with the solar wind and magnetospheric heavy ions and also by
the bremsstrahlung emission via energetic electrons \cite{gla02,bra04,bra07a}. 
A typical X-ray luminosity of the auroral emission is $\sim$10$^{16}$ erg s$^{-1}$.
The low-latitude atmosphere also exhibits scattered radiation of solar 
X-rays with a typical luminosity of $\sim3\times10^{15}$ erg s$^{-1}$\cite{bra07b}. 
Besides Jupiter itself, faint soft ($<$ 1 keV) X-rays from the Io Plasma 
Torus (IPT) have been detected ($\sim1\times10^{14}$ erg s$^{-1}$, Elsner et al. 2002). 
It may arise from bremsstrahlung emission by warm electrons with energies 
of few hundred to few thousand eV. 
Although we can expect hard X-rays ($>$ 1 keV) from energetic keV to MeV 
electrons in the Jupiter's radiation belts, there has been no report on
such an emission because past instruments lacked a necessary sensitivity 
to detect the extended hard X-ray emission.
This indication motivated us to perform a decisive analysis using a deep 
160 ks observation data of Jupiter with {\it Suzaku}.

\section{Observations}
\label{sec:intro}

{\it Suzaku} \cite{mit07} observed Jupiter on 2006 February 24-28 with 
the X-ray Imaging Spectrometer (XIS) \cite{koy07}. The XIS consists of
three front-illuminated (FI) CCDs (XIS0, 2 and 3) and one back-illuminated
(BI) CCD (XIS1). 

Due to the low-earth orbit of {\it Suzaku} and the large effective area, 
the XIS-FI has the lowest particle background among all X-ray CCDs in 
currently available X-ray observatories. 
Figure 5 in Mitsuda et al. (2007) summarizes the background 
normalized by the effective area and the field of view, which is 
a good measure of sensitivity determined by background for 
spatially extended emission. 
The background of the XIS FI is the lowest among X-ray CCDs onboard
{\it Suzaku}, {\it Chandra} and {\it XMM-Newton}. It is 2 and 5 times 
lower than the {\it XMM-Newton} MOS and pn, and {\it Chandra} ACIS-I 
in 2--5 keV, respectively. 
For this reason, we use only the XIS data in this paper.

During the observations, the spacecraft was repointed four times to allow 
for the planet's motion (0.7$\sim$1.8 arcmin per day). The XIS was operated 
in the normal mode. 
The data reduction was performed on the version 2.0.6.3 screened data 
provided by the Suzaku processing facility, using the HEAsoft analysis 
package ver 6.3.1. The net exposure of each FI and BI chip was 159 ks.
For spectral fits, we generated response matrices and auxiliary files
with {\tt xisrmfgen} and {\tt xissimarfgen}. 

Although an optical loading from Jupiter (visual magnitude $m_{\rm V}=-2$
during Suzaku observations) often becomes a problem in X-ray observations 
\cite{els02}, we concluded that it is negligible in this case, because 
optical loading effects such as degradation of the energy resolution, 
changes in the energy scales and/or the detection efficiency were not 
seen in XIS spectra. 
For confirmation, we estimated the optical blocking power of the XIS 
compared to CCDs onboard {\it XMM-Newton} \cite{lum00}.
We found that the {\it Suzaku} XIS has almost the same blocking power 
as the {\it XMM-Newton} MOS with a thick filter, with which the 
observation of Jupiter was successfully conducted \cite{bra04}.

\section{Extended X-ray Emission}
\label{sec:ext}

\subsection{Imaging analysis}
\label{sec:sec:image}

To search for any emission from Jupiter, we have to remove {\it Suzaku}'s orbital 
motion and transform the data into Jupiter's comoving frame using an ephemeris obtained
from the Jet Propulsion Laboratory (JPL). Before conducting this procedure, we 
checked X-ray images without these corrections. 
We created two-band mosaic images using XIS1 (BI) and XIS0$+$2$+$3 (FI) data 
as shown in figure \ref{fig:xisimg-obs}. We corrected the data for positional 
difference in exposure times by using the exposure map generator {\tt xisexpmapgen}.
An extended emission is detected in both the 0.2--1 keV and 1--5 keV bands along 
the path of Jupiter. 

Beside the extended emission, we found signatures of emission from point 
sources near Jupiter in the 1--5 keV band.
Since we found no known point sources in the X-ray source catalogue compiled by 
NASA$\footnote{http://heasarc.gsfc.nasa.gov/cgi-bin/Tools/high\_energy\_source/high\_energy\_source.pl}$,
we decided to detect sources from the {\it Suzaku} image.
We utilized the wavelet function program {\tt wavdetect} in the CIAO package
$\footnote{http://cxc.harvard.edu/ciao/}$. We created four 1--5 keV images
with different bin sizes (1, 2, 4, and 8) and ran the point search program
by setting the significant threshold above 1$\times10^{-6}$, which 
corresponds to one spurious source in a 1024$\times$1024 pixel map 
(XIS field of view). 
We summarized source lists and removed possible spurious detections by 
visual inspection.

The obtained source positions are shown in figure \ref{fig:xisimg-obs} (b) 
(green circles with red lines).
Twenty sources are detected with the 1--5 keV X-ray flux of
$1\sim3\times10^{-14}$ erg s$^{-1}$ cm$^{-2}$. 
Here we converted a XIS count rate into flux, assuming a power-law spectrum 
with an photon index of 1.5 and an average absorption column toward this 
field$\footnote{http://heasarc.gsfc.nasa.gov/cgi-bin/Tools/w3nh/w3nh.pl}$,
which represents a typical spectrum of the background active galactic nucleus.
This number is consistent with the canonical cosmic X-ray background model 
\cite{gia01}, which predicts 27 background sources above the detection limit. 
Although the source 
list contains marginally detected point sources, we decided to remove X-ray 
photons from the 1--5 keV image analysis afterward, for safety. 
We excluded circular regions with a diameter of 2 arcmin centered at 
individual sources. The diameter is equal to the beam size or the half
power diameter of the XIS $\footnote{http://www.astro.isas.jaxa.jp/suzaku/doc/suzaku\_td}$.
Since all the sources are faint, contamination from the sources is thus
negligible.
Because there are no significant point sources around Jupiter in the 
0.2--1 keV image (fig. \ref{fig:xisimg-obs} a), we conducted this 
exclusion process only for the 1--5 keV image.

We then corrected the two-band images for {\it Suzaku}'s orbital
motion and Jupiter's ephemeris. 
In case of the 1--5 keV image, to take into account the excluded 
regions, we created the exposure map for each 
observation and excluded the point source regions.
Then we transformed the exposure map considering the Jupiter's orbit
and divided the orbital motion corrected image by the 
corrected map.
In case of the 0.2--1 keV image, we made the same analysis
except for excluding the point source regions.

The results obtained are shown in figure \ref{fig:xisimg-sta}.
We detected a significant hard X-ray emission extended over 
$\sim16\times8$ $R_{\rm j}$ which is positionally associated 
with the Jupiter's radiation belts, and Io's orbit.
Because Io's orbit coincides with the IPT,
the emission is also spatially associated with the IPT.
The X-ray luminosity of the extended emission arises 
from (3.6$\pm$0.4)$\times$10$^{15}$ erg s$^{-1}$ and 
(3.3$\pm$0.5)$\times$10$^{15}$ erg s$^{-1}$ in 0.2--1 
and 1--5 keV, respectively. Here and below the uncertainties 
are 68 \% confidence intervals. We assumed a distance 
to Jupiter of 5.0 AU on 24-28 Feb 2006.
An estimated contamination of the point sources outside
the excluded regions to the extended emission is $\sim$20 \%.
Hence, most of the emission is due to the extended component.

To check the distribution of the emission, 
we created projection profiles along the 
horizontal axis as shown in figure \ref{fig:xisimg-sta-cr}.
The hard X-ray emission is significantly extended more 
than the point spread function (green line).
We found that the profile can be roughly represented by 
a simulated model using {\tt xissimarfgen}, which is
a sum of a point source and a elliptical uniform emission 
with radii of 4 and 1.3 arcmin (12 and 4 $R_{\rm j}$). 
The emission is thus extended over a wide region ($>6$ $R_{\rm j}$). 
On the other hand, the soft emission is marginally consistent 
with the point source. There is an excess on the right side 
of the soft X-ray peak, which suggests emission from the IPT 
that has been reported by Elsner et al. (2002).

The hard X-rays are hence unlikely to originate only from 
a single point source or Jupiter. 
Furthermore, an artificial effect on the extended morphology 
due to the analysis can be excluded because of the different 
morphology from the soft X-ray image.
The emission thus seems to contain truly diffuse component 
related to Jovian magnetospheric processes.

We note that the faintness of the emission should have hindered
its detection in the past observations.
Considering the angular resolution of {\it XMM-Newton}, 
its surface brightness will be 1/30$\sim$1/40 of Jupiter's aurora, 
From figure 5 in Branduardi-Raymont et al. (2007a), we can 
roughly estimate the CCD background level as 1/10 of the 
auroral emission.
Therefore the detection of the diffuse emission shuold be 
quite difficult. If the emission is related to Jovian 
radiation belts, there can be also time variation (see \S 4).

\subsection{Spectral analysis}
\label{sec:sec:spec}

To investigate characteristics of the emission,
we analyzed its X-ray spectrum by extracting photons
around the extended emission region.
For simplicity, we excluded the point sources from the 
event files without orbital motion corrections and 
extracted events from a circular region with a radius 
of 3 arcmin centered at the path of Jupiter. 
The surrounding region with an outer radius of 6 arcmin 
after subtracting the point sources was utilized as a background.
Consequently, this extended emission region includes both Jupiter and 
the extended emission, because the limited angular resolution 
hindered us to spatially separate these two components.

Figure \ref{fig:xisspec} shows the spectrum obtained.
It consists of a flat continuum extending up to 5 keV and 
signs of lines around 0.25 keV and 0.56 keV. Hence, we fitted 
the spectrum with a power-law plus two Gaussian model. 
The best-fit model represents data well with $\chi^2/\nu\sim$0.5.
No absorbing column was required, as expected from the source's 
proximity.
The obtained photon index was 1.4$\pm$0.2, while the central 
energies of the two Gaussian were 0.24$\pm$0.1 keV and 
0.56$\pm$0.1 keV. Fluxes for the 0.24 keV and 0.56 keV lines
were (3.9$\pm$0.6)$\times$10$^{-5}$ and 
(2.0$\pm$0.5)$\times$10$^{-5}$ photons cm$^{-2}$ s$^{-1}$, 
respectively. 

Characteristics of the two lines such as a central energy and a flux 
were similar to those of the Jupiter's auroral emission observed
with {\it XMM-Newton} in Nov 2003 (see table 3 in Branduardi-Raymont et al. 2007a). 
The two lines at 0.24 keV and 0.56 keV are presumably a line complex 
from C$^{5+}$ and ionized Mg, Si, S and a K$_\alpha$ emission from O$^{6+}$, 
respectively. Such line emissions are distinct features of the charge 
exchange reaction \cite{cra03}. Taken together with the fact that the 0.2--1 keV X-rays
arise from a compact region near Jupiter, most of soft X-rays likely 
originate from Jupiter's aurora via the charge exchange. 

In contrast, the 1--5 keV X-ray emission was represented by a simple 
flat continuum with a photon index of 1.4. Since the peak of the 
X-rays coincides with Jupiter (fig. \ref{fig:xisimg-sta} b), some 
fractions of the emission can arise from Jupiter. 

The power-law spectrum was in fact seen in the past {\it XMM-Newton} 
observations of Jupiter's aurora \cite{bra04}.
For comparison, we plotted auroral emission models in three 
past {\it XMM-Newton} observations (tables 2 and 3 in 
Branduardi-Raymont et al. 2007a) in figure \ref{fig:xisspec}.
The 1--5 keV fluxes of these models are smaller than that 
of the {\it Suzaku} best-fit model by a factor of 2$\sim$3. 
This is consistent with an independent estimation from the projection 
profile that most of the extended emission comes from diffuse emission
other than Jupiter.

\section{Discussion}
\label{sec:diss}

\begin{figure}
\begin{center}
\includegraphics[width=160mm]{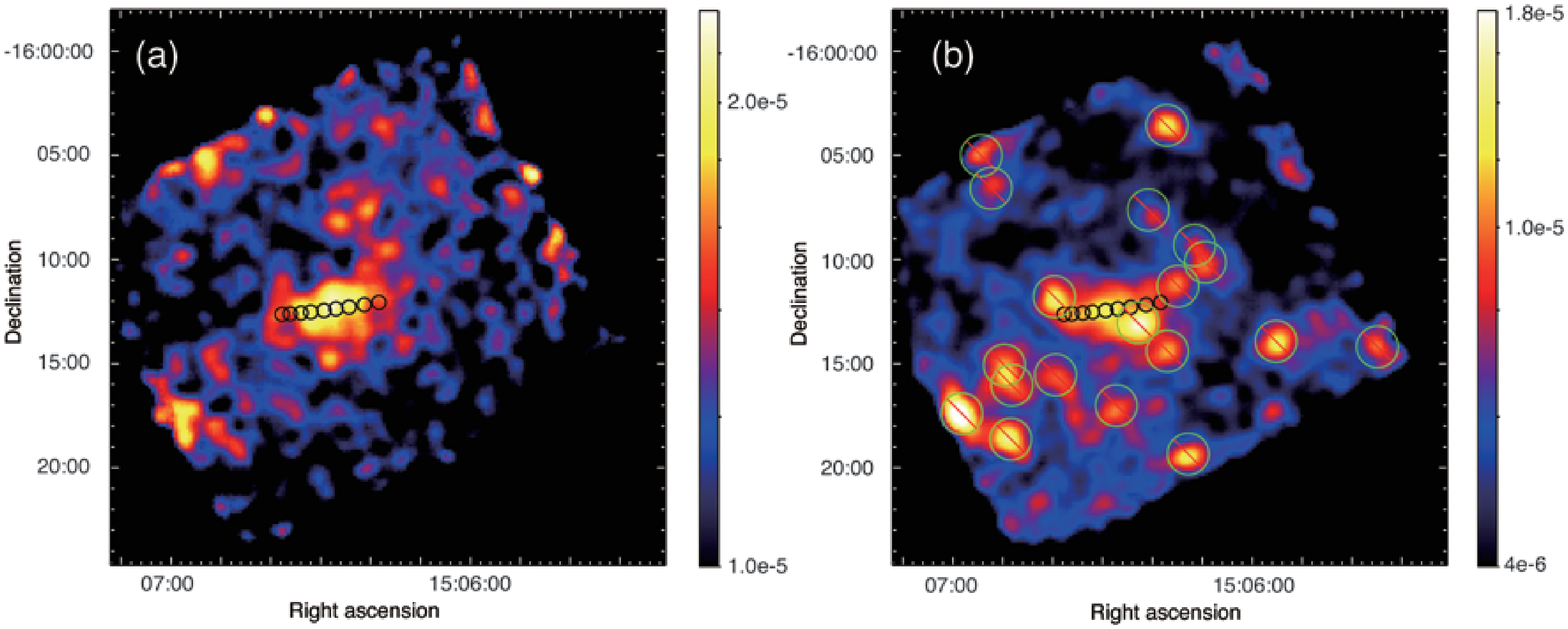}
\caption{{\it Suzaku} XIS mosaic images of the vicinity of Jupiter in the (a) 0.2--1 keV (BI) 
and (b) 1--5 keV (FI) bands, displayed on the J2000.0 coordinates. Exposures are corrected
and the count unit is counts s$^{-1}$ binned pixel$^{-1}$. For clarity, images are binned 
by a factor of 8 and smoothed by a Gaussian of $\sigma=$ 5 pixels. Black circles show the size and position of 
Jupiter during the observations. Considering known pointing uncertainty of {\it Suzaku} 
\cite{uch08}, positions of the circles are slightly shifted by $+$25 and $-25$ arcsec in the 
Ra and Dec directions, respectively, in order that the extended emission coincides with the 
Jupiter's path. Green circles with red lines mark excluded point source regions when 
we create the 1--5 keV image in comoving frame (see text).}
\label{fig:xisimg-obs}
\end{center}
\end{figure}

\begin{figure}
\begin{center}
\includegraphics[width=160mm]{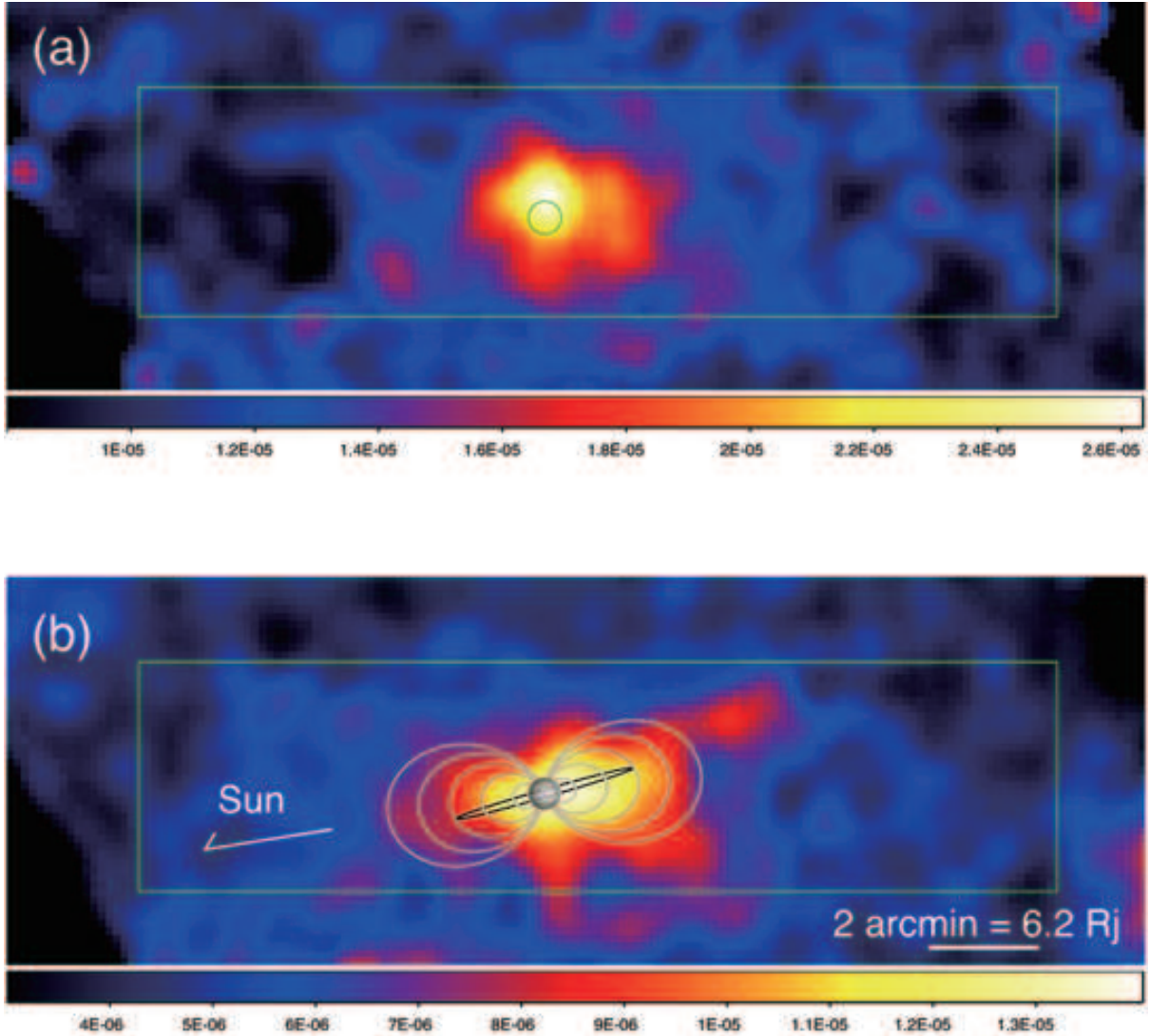}
\caption{{\it Suzaku} XIS images after correcting for the satellite's orbital motion
and Jupiter's ephemeris in the (a) 0.2--1 keV (BI) and (b) 1--5 keV (FI) bands. 
The images are binned and smoothed in the same way as figure \ref{fig:xisimg-obs}.
In the panel (a), a circle indicates the expected position and size of Jupiter 
whose diameter is 39 arcsec. 
In the panel (b), grey lines indicate the equatorial crossing of magnetic field lines 
at 2, 4, 6, and 8 Jovian radius ($R_{\rm j}$). 
A black line is the path traced by Io. A photograph of 
Jupiter by {\it Cassini} taken from the JPL web site is overlaid. 
An arrow indicates the direction of the Sun. 
Boxes are used to obtain the projection profiles 
in figure \ref{fig:xisimg-sta-cr}.
}
\label{fig:xisimg-sta}
\end{center}
\end{figure}

\begin{figure}
\begin{center}
\includegraphics[width=160mm]{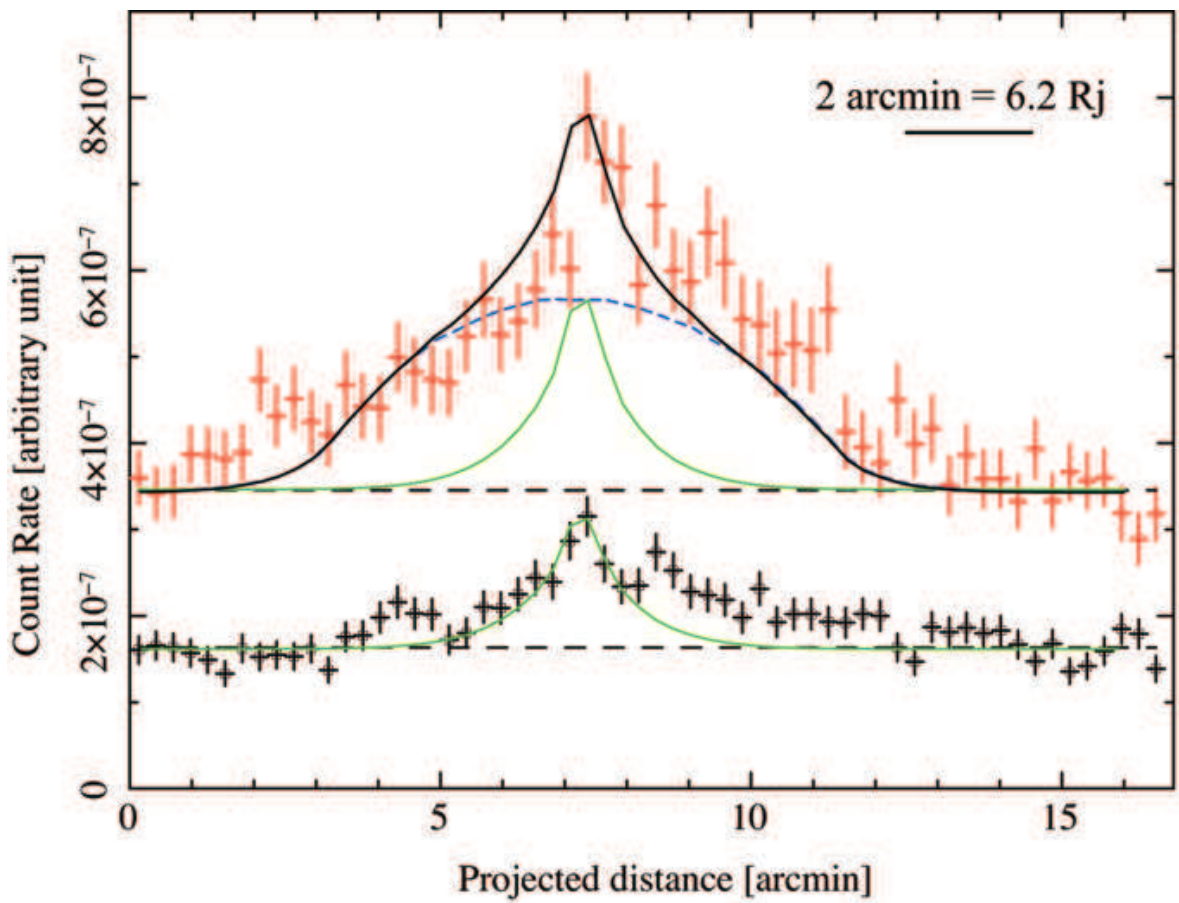}
\caption{Projection profiles along the horizontal axis, 
extracted from box regions in figure \ref{fig:xisimg-sta}.
Black and red points show the 0.2--1 and 1--5 keV data, respectively. 
Errors are 1$\sigma$ statistical ones.
Green lines are a simplified model of Jupiter's X-rays (42 arcsec diameter) 
in the two energy bands, whose normalization and offset are tuned to coincide 
with the profile. 
The point spread function is almost the same at different energies 
\cite{ser07}.
A dashed blue line indicates the expected profile of an uniform emission
extended over an elliptical region with radii of 4 and 1.3 arcmin.
A solid black line is the sum of the dashed green and blue lines.
Dashed black lines are background levels estimated by fitting the data 
above the projected distance $>$14 arcmin with a constant.
}
\label{fig:xisimg-sta-cr}
\end{center}
\end{figure}

We have discovered extended hard X-ray emission around Jupiter. Although
its spatial association with Jupiter strongly suggests a diffuse 
component related to Jovian magnetospheric processes, we first have 
to examine serendipitous background sources.
Because we corrected the image for the Jupiter's orbital motion,
any faint background source on the Jupiter's path, that is missed 
in our point source identification procedure, can be seen as extended 
emission in the orbital motion corrected image.
Considering the almost symmetric projection profile (fig. 
\ref{fig:xisimg-sta-cr}) centered at Jupiter, the spatial distribution
of the background sources should be symmetric about Jupiter.
The most plausible case is that the emission is composed of Jupiter 
itself and a single background point source near the center of the 
Jupiter's orbit. 
To test this hypothesis, we excluded the central part of the 
Jupiter's orbit with a circle region in the same way as figure \ref{fig:xisimg-obs} (b),
and created the orbital motion corrected image and projection
profile in 1--5 keV.
We still observed a significant extended emission around Jupiter,
although the total count rate decreased by $\sim$15 \%. Hence, a single 
background source cannot fully explain the observed emission. 
Since multiple point sources with symmetric spatial distribution 
against Jupiter are less probable, below we discuss the origin of 
the emission, considering that the emission is truly diffuse.
 
The flat power-law continuum of the emission suggests a non-thermal mechanism. 
There are three candidates for the emission mechanism; synchrotron emission, 
bremsstrahlung emission, and inverse-Compton scattering.
The synchrotron interpretation can be easily rejected 
in terms of the required electron energy. From the X-ray image, we can 
observe the hard X-rays at the Io's orbital plane (a distance from Jupiter, 
$r=$5.9 $R_{\rm j}$) or even more distant area.
At $r\sim$6 $R_{\rm j}$, the magnetic field of the Jovian magnetosphere
is quite weak on the order of 0.01 G. With such a weak magnetic field, 
if we want to generate synchrotron emission in the X-ray wavelength range, 
TeV electrons are necessary. This energy is more than 10$^{4}$ times 
higher than the observed maximum electron energy in the radiation belts 
($\sim$50 MeV) \cite{bol02}. 
Even if TeV electrons exist, 
their Larmor radius ($3\times10^{11}$ cm for 1 TeV electron at 
$r=$6 $R_{\rm j}=4\times10^{10}$ cm) 
will be too large, so that TeV electrons are not trapped
in the magnetic field line and will escape from the 
Jupiter's radiation belts.

\begin{figure}
\begin{center}
\includegraphics[width=160mm]{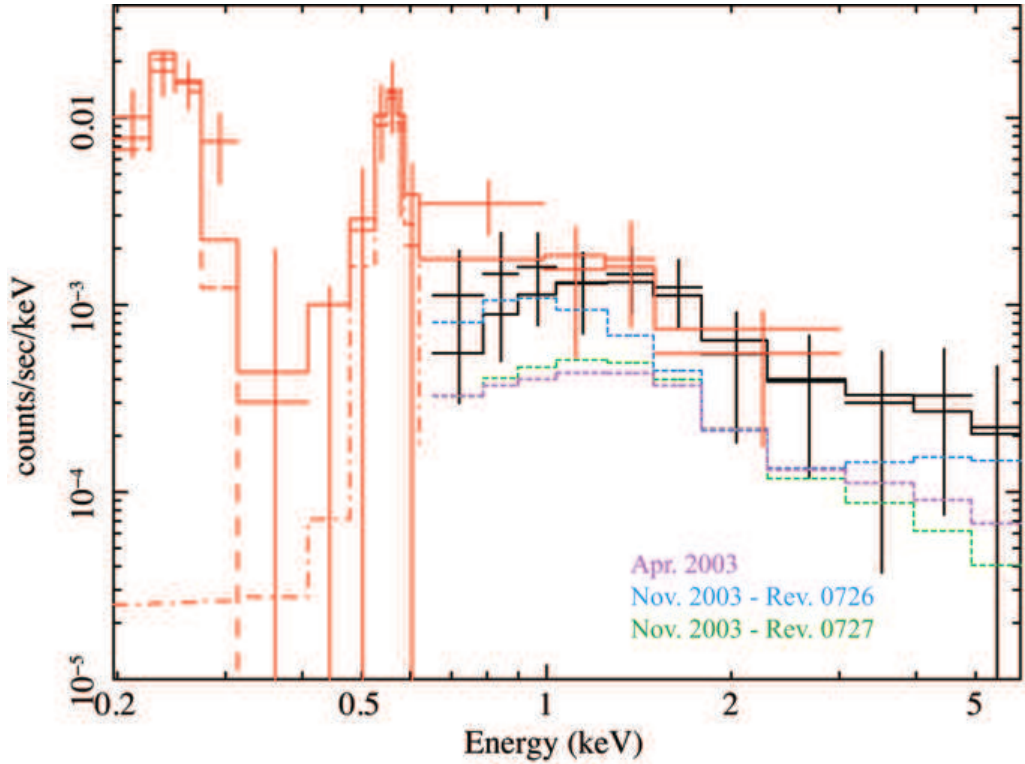}
\caption{Background subtracted BI (red) and FI (black) spectrum of the extended emission region,
compared with the best-fit power-law plus two Gaussian models (solid line). Two Gaussians are 
shown in dashed and dash-dotted lines. 
Purple, green and blue dashed lines plotted for the FI spectrum are Jupiter's auroral continuum 
emission models in {\it XMM-Newton} observations (Branduardi-Raymont et al. 2007a). 
}
\label{fig:xisspec}
\end{center}
\end{figure}

The bremsstrahlung emission is possible in terms of the electron energy, 
because we need only keV electrons which have been observed in the past 
in-situ measurements \cite{bag04}. 
Such electrons will interact in an ambient medium and produce 
not only continuum but also X-ray lines by collisional excitation 
and inner-shell ionization (see \S3.1 in Tatischeff 2002).
From in-situ measurements, the plasma at $r\sim$6 $R_{\rm j}$ is 
known to be mainly composed of heavy ions such as S$^{+}$ and O$^{+}$ 
(see fig. 23.2 in Bagenal et al. 2004). 
We estimated an equivalent width of S K$_{\alpha}$ line (2.3 keV). 
We used cross sections for K-shell ionization 
by electron impact from Santos et al. (2003) and  the bremsstrahlung 
loss rate of electrons given by Skibo et al. (1996), to estimate 
the line and continuum intensities, respectively.
The calculated equivalent width of the S K$_{\alpha}$ line was as large 
as $\sim$10 keV, which should be observed in the XIS spectrum with the 
good energy resolution ($\sim$130 eV at 6 keV, Koyama et al. 2007).
As shown in figure \ref{fig:xisspec}, no such line exists around 2 keV.
Hence, the bremsstrahlung interpretation can be rejected.

The remaining possibility is inverse-Compton scattering.
We consider the scattering of solar light by ultra-relativistic 
electrons in the radiation belts. 
In this process, the scattered photon energy will be
\begin{equation}
\sim8~{\rm keV} \left( E_{\rm ph}/1.4~ {\rm eV} \right) ~ \left( E_{\rm e} / 50~ {\rm MeV} \right)^2, 
\end{equation}
where $E_{\rm ph}$ is an energy of solar optical photons 
and $E_{\rm e}$ is that of relativistic electrons \cite{ryb79}. 
Considering the maximum observed electron energy of 50 MeV in 
the inner radiation belts \cite{bol02}, the inverse-Compton 
scattering is possible in terms of the electron energy.
Although 50 MeV electrons have not been observationally well 
confirmed at outer regions ($r>$ several $R_{\rm j}$), 
there may exist such ultra-relativistic electrons
in these regions as expected from the empirical model 
(e.g., Divine \& Garret 1983).
Also, since this mechanism does not need the emission line, 
the observed spectral shape can be safely explained.

Assuming the inverse Compton scattering, we can estimate the index 
of the electron number density spectrum as $\sim2$, from the photon 
index of the X-ray spectrum \cite{ryb79}. This is consistent 
with the Divine-Garrett (DG) empirical charged particle model of 
Jovian magnetosphere \cite{div83} ($\sim3$ at 6 $R_{\rm j}$) 
and another model based on the {\it Galileo} data \cite{gar03} 
($\sim3$ at 8 $R_{\rm j}$).

We then proceeded to estimate the necessary electron number 
density. For simplicity, we assumed that the emission region is an 
oblate spheroid with radii of 8, 8, and 4 $R_{\rm j}$, and the monotonic 
electron energy of 50 MeV. 
We here considered anisotropic angular distribution of the 
inverse-Compton scattering \cite{bru01}. During the {\it Suzaku} 
observation, the Sun-Jupiter-Earth angle was 10$^\circ$. Because of 
the larger cross section of the back scattering, the inverse-Compton 
flux will increase by a factor of $\sim$3.
The estimated average electron number density to explain 80\% of 
the hard X-rays was $\sim$0.005 cm$^{-3}$.

For comparison, we utilized the DG empirical model,
which provides a theoretical electron spectrum as a function
of the distance from Jupiter.
The DG model is based on the data taken with {\it Pioneer} 
and {\it Voyager}. To date, it is the best model to estimate the 
MeV electron distribution in the inner radiation belts
($<8$ $R_{\rm j}$).
We found that the $>$20 MeV electron number density is only
$\sim$0.0007 and 0.0001 cm$^{-3}$ at 4 and 6 $R_{\rm j}$, 
respectively. The density will decrease as the distance increases. 
Hence, there is a factor of 7$\sim$50 discrepancy. 
Although the situation can be somewhat relaxed by an uncertainty 
of the DG model \cite{bol01} and/or time variations 
of the trapped MeV electrons \cite{miy99,bol02,san08},
this interpretation needs further investigations.

In conclusion, further X-ray, radio and in-situ observations 
are necessary to understand the origin of this diffuse hard 
X-ray emission, to identify the emission region with the 
Jupiter's magnetosphere, the IPT and/or else, and to clearly 
distinguish the emission from background point sources.

\bibstyle{natbib}

\clearpage

\clearpage

\clearpage

\clearpage

\end{document}